%% file: moonshine_paper.tex
\documentclass[nohyperref]{article}

\usepackage{microtype}
\usepackage{graphicx}
\usepackage{subfigure}
\usepackage{booktabs} 
\usepackage[dvipsnames]{xcolor}
\usepackage{soul}
\usepackage{dblfloatfix}
\usepackage{import}

\usepackage{hyperref}
\usepackage{multirow}
\usepackage{tabularx}
\usepackage{colortbl}
\usepackage{hhline}

\newcolumntype{C}{>{\centering\arraybackslash}X}

\usepackage[accepted]{icml2022}

\usepackage{amsmath}
\usepackage{amssymb}
\usepackage{mathtools}
\usepackage{amsthm}

\theoremstyle{plain}

\theoremstyle{definition}

\theoremstyle{remark}

\usepackage[textsize=tiny]{todonotes}

\icmltitlerunning{Moonshine: Speech Recognition for Live Transcription and
  Voice Commands}
\begin{document}

\twocolumn[
  \icmltitle{Moonshine: Speech Recognition for Live Transcription and
    Voice Commands}



\icmlsetsymbol{equal}{*}

\begin{icmlauthorlist}
\icmlauthor{Nat Jeffries}{}
\icmlauthor{Evan King}{}
\icmlauthor{Manjunath Kudlur}{}
\icmlauthor{Guy Nicholson}{}
\icmlauthor{James Wang}{}
\icmlauthor{Pete Warden}{}
\end{icmlauthorlist}

\begin{center}
\href{https://usefulsensors.com}{Useful Sensors}
\end{center}
\icmlcorrespondingauthor{Manjunath Kudlur}{keveman@usefulsensors.com}

\icmlkeywords{Speech Recognition, Machine Learning}

\vskip 0.3in
]



\printAffiliationsAndNotice{} 

\begin{abstract}

This paper introduces Moonshine, a family of speech recognition models optimized
for live transcription and voice command processing. Moonshine is based on an
encoder-decoder transformer architecture and employs Rotary Position Embedding
(RoPE) instead of traditional absolute position embeddings. The model is trained
on speech segments of various lengths, but without using zero-padding, leading
to greater efficiency for the encoder during inference time. When benchmarked
against OpenAI's Whisper \verb|tiny.en|, Moonshine Tiny demonstrates a 5x reduction in
compute requirements for transcribing a 10-second speech segment while incurring
no increase in word error rates across standard evaluation datasets. These
results highlight Moonshine's potential for real-time and resource-constrained
applications.

\end{abstract}

\section{Introduction}
\label{sec:introduction}
\import{sections}{introduction.tex}

\section{Issues with fixed sequence length encoder}
\label{sec:fixed-length-encoder-issues}
\import{sections}{motivation.tex}

\section{Approach}
\label{sec:approach}
\import{sections}{approach.tex}

\section{Evaluations}
\label{sec:evaluation}
\import{sections}{evaluation.tex}

\section{Discussion \& Conclusion}
\label{sec:conclusion}
\import{sections}{conclusion.tex}

\section{Acknowledgements}
\import{sections}{ack.tex}

\nocite{*}
\bibliography{moonshine_paper}
\bibliographystyle{icml2022}

\end{document}

%% file: sections/introduction.tex
Real-time automatic speech recognition (ASR) is essential for many applications, including live transcription during presentations, accessibility tools for individuals with hearing impairments, and voice command processing for conversational interfaces in smart devices and wearables. These applications often run directly on low-cost hardware, where strict resource constraints and a lack of internet connectivity introduce unique technical challenges that are not present in other ASR domains.

The introduction of OpenAI's Whisper has led to a significant improvement in the accuracy of general-purpose ASR systems~\cite{radford2022robustspeechrecognitionlargescale}. However, one of the primary challenges in applications of \emph{on-device} ASR is the need to minimize latency---the time delay between a spoken utterance and the corresponding text appearing, e.g., on a live transcription display---without losing accuracy. During our development of one such application---a Caption Box built to provide rapid, accurate, and private offline transcription of English speech---we found that current models are ill-suited to the task. When deployed on a low-cost ARM-based processor, we noticed that even the smallest Whisper \verb|tiny.en| model had a firm lower latency bound of 500 milliseconds, irrespective of the duration of audio. User feedback indicated that this level of latency failed to provide a smooth and responsive user experience, which motivated us to investigate deeper.

Whisper, which is based on an encoder-decoder transformer architecture, processes fixed-length audio sequences in 30-second chunks regardless of the actual speech content of the audio. This means that shorter audio sequences require padding with zeros to meet the length requirement, resulting in a constant computational overhead in the encoder. While the decoder's processing time is proportional to the length of the utterance, the constant overhead of fixed-length encoding places a firm lower bound to latency---e.g., the 500 milliseconds we identified during our own testing with the Caption Box. Our initial attempts to remove this bottleneck involved fine-tuning and distilling~\footnote{Hence the name Moonshine. We progressed to not use distillation eventually, but the name stuck.} Whisper models to handle variable-length sequences in the encoder, utilizing open audio datasets. However, these open datasets proved insufficient to surpass Whisper's Word Error Rate (WER). Recognizing the limits of available data and the opportunity to leverage recent advancements in model architectures, we opted to develop and train new models from scratch.

\begin{figure}[]
  \centering
  \includegraphics[width=\columnwidth]{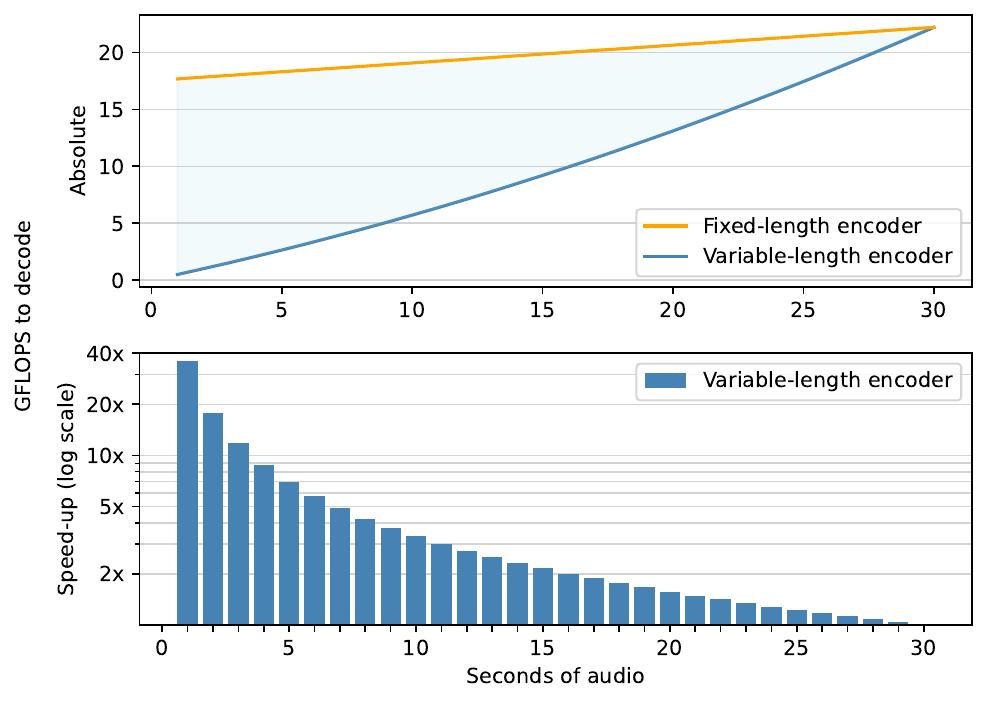}
  \vspace{-2ex}
  \caption{GFLOPS required by Whisper \texttt{tiny.en} for decoding as a function of input audio duration. OpenAI's Whisper models use a fixed-length encoder that requires any audio less than 30 seconds to be zero-padded; testing with a hypothetical \emph{variable}-length encoder shows that we can attain significant speed-ups by removing this requirement.}
  \label{fig:decoding_gflops}
\end{figure}

Our first task was to better quantify the bottleneck imposed by Whisper's fixed-length encoder. In an empirical test, we compared the GFLOPS (i.e., billion floating-point operations) needed for processing 30 seconds of zero-padded audio to the GFLOPS that would be required to instead process only a $< 30$ second subset of the same audio---i.e., we measured the potential for speed-ups if Whisper instead had a \emph{variable}-length encoder. Our results (Figure~\ref{fig:decoding_gflops}) showed potential for a 35x speed-up in the best case, and nearly a 5x speed-up overall. Clearly, improvements to Whisper's architecture could yield lightweight, low-latency speech-to-text models that are better-suited to resource-constrained applications.

Toward tackling the challenges faced by on-device, low-latency ASR applications, this paper introduces the Moonshine family of ASR models. Moonshine models are designed to match Whisper's accuracy while optimizing computational efficiency by eliminating zero-padding requirements, instead scaling processing demands proportionally to audio input length. We describe Moonshine's architecture and training process, and provide a detailed analysis of its performance benefits in comparison with Whisper models. Section~\ref{sec:fixed-length-encoder-issues} motivates our development of Moonshine by quantifying the WER when adapting Whisper for variable-length audio. Section~\ref{sec:approach} describes the Moonshine architecture, dataset preparation, and training process, while Section~\ref{sec:evaluation} provides an evaluation of the results on standard speech recognition datasets. Section~\ref{sec:conclusion} concludes.

%% file: sections/motivation.tex
\begin{figure*}[t!]
  \centering
  \includegraphics[width=\textwidth]{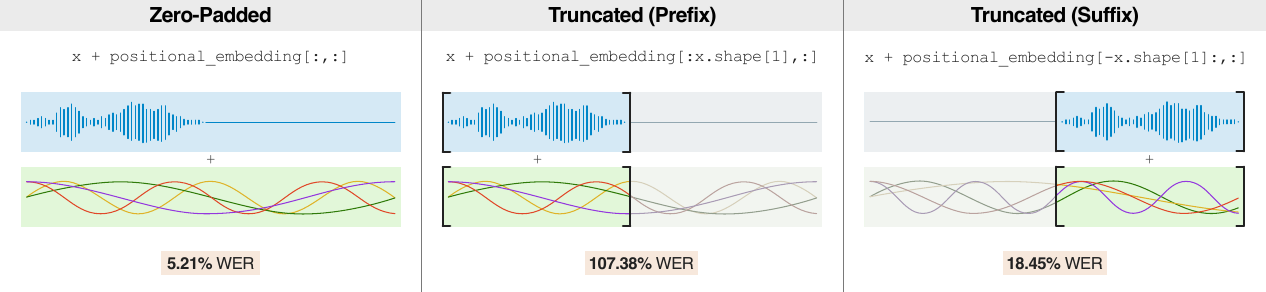}
  \caption{Variations of absolute position embeddings usage and the corresponding WER produced by Whisper \texttt{tiny.en} (\texttt{test.clean} split of the Librispeech dataset). Simply adapting Whisper's inference code to avoid encoding of fixed-length audio (as in the center and right-most columns) introduces significant increases in WER, motivating our development of new models with variable-length encoding.}
  \label{fig:encoding_segments}
\end{figure*}

Whisper models use absolute position embeddings in both the encoder and decoder. The encoder in particular uses sinusoidal position embeddings of shape \verb|[1500, dim]|, where \verb|dim| is the transformer encoder's dimension. The standard inference code provided by OpenAI zero-pads the input audio to an exact 30-second segment before passing it down to the model. Audio feature generation and 2 convolution layers convert this audio segment to a \verb|[1500, dim]| vector sequence that gets added to the sinusoidal position embedding. The left-most column of Figure~\ref{fig:encoding_segments} illustrates this mechanism.

Requiring an exact 30-second segment causes the encoder to incur a fixed computation cost, no matter the duration of actual speech in the audio segment. However, since Whisper's model architecture is an off-the-shelf, variable-length sequence-to-sequence transformer, zero-padding is not strictly necessary. An audio segment that is shorter than 30 seconds can indeed be processed through the model's frontend to yield a \verb|[seq_length, dim]| vector sequence where \verb|seq_length <= 1500|. 

There are several options for adding the position embedding to a variable-length audio input. We compared the default implementation in OpenAI's Whisper inference code (i.e., using zero-padding to form 30-second segments) with two other options, as illustrated in the center and right-most columns of Figure~\ref{fig:encoding_segments}---namely, using the prefix and the suffix of the position embedding. We computed the WER for the \verb|tiny.en| model using these two options on the \verb|test.clean| split of the Librispeech dataset. This split has 2620 examples, 9 of which are more than 30 seconds in length. We ignore these 9 examples in this evaluation. We use a beam size of 5, so as to not suffer the pitfalls of greedy decoding such as repetitions. With the standard inference code provided by OpenAI, we get a WER of 5.21\%. When the input is \emph{not} zero-padded, however, we get worse results. When using only a prefix of the position embedding, the \verb|tiny.en| model has a WER of 107.38\%. We notice a lot of repetitions towards the end of the transcription, even with the beam search. When instead using the suffix of the position embedding, the accuracy degradation is not as dramatic, producing a WER of 18.45\%---still more than three times worse than the baseline.

These results indicate that the OpenAI Whisper models are trained with examples that are all exactly 30 seconds long. The models indeed produce high quality transcriptions for long-form audio. However, the fixed computation budget needed for their encoder makes them inefficient for low-latency applications, such as live transcription. Since the audio segments in these applications tend to be short and/or vary in length, our effort to train models optimized for variable-length sequences is well-motivated.

%% file: sections/approach.tex
\begin{figure*}[]
  \centering
  \includegraphics[width=0.66\textwidth]{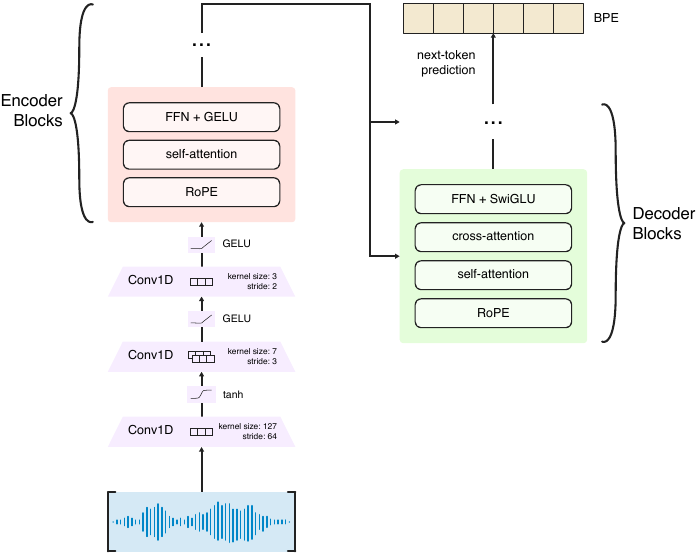}
  \caption{Moonshine's model architecture.}
  \label{fig:model-arch}
\end{figure*}

This section describes Moonshine's model architecture before detailing our data collection, preprocessing, and model training process.

\subsection{Model architecture}
Moonshine is an off-the-shelf encoder-decoder Transformer \cite{vaswani2017transformer} model. The input is an audio signal sampled at 16,000 Hz. We do not use any hand-engineering (e.g., Mel spectrograms) to extract audio features. Instead, the input is processed by a short stem of 3 convolution layers with strides 64, 3, and 2 respectively. The filter widths, number of channels and activation functions are shown in Figure~\ref{fig:model-arch}. These set of strides compress the input by a factor of 384x. Note that in the Whisper model, the input is compressed by 320x---firstly 160x in the Mel spectogram computation, then by a further 2x in the convolution stem. We use Rotary Position Embeddings (RoPE)~\cite{su2023roformerenhancedtransformerrotary} at each layer of the encoder and decoder. Table~\ref{tab:model-shapes} compares Moonshine's architecture with OpenAI's Whisper models.

{
  \renewcommand{\arraystretch}{1.25}
  \begin{table*}[ht]
    \small
    \centering
    \begin{tabular}{|c|c|c||c|c|}

    \hline
    \textbf{Parameter} & \textbf{Moonshine Tiny} & \textbf{Whisper \texttt{tiny.en}} &
    \textbf{Moonshine Base} & \textbf{Whisper \texttt{base.en}} \\
    \hline \hline
    Dimension & 288 & 384 & 416 & 512 \\
    \hline
    Encoder layers & 6 & 4 & 8 & 6 \\
    \hline
    Decoder layers & 6 & 4 & 8 & 6\\
    \hline
    Attention heads & 8 & 6 & \multicolumn{2}{|c|}{8}\\
    \hline
    Encoder FFN activation & \multicolumn{4}{|c|}{GELU} \\
    \hline
    Decoder FFN activation & SwiGLU & GELU & SwiGLU & GELU \\
    \hline
    Parameters in Millions & 27.1 & 37.8 & 61.5 & 72.6 \\
    \hline
    FLOPs normalized to Whisper \texttt{tiny.en} & 0.7x & 1.0x & 1.6x & 2.3x \\
    \hline
    \end{tabular}
    \caption{Model shapes and GFLOPs}
    \label{tab:model-shapes}
  \end{table*}
}

We use the same byte-level BPE text tokenizer used in Llama 1 and 2~\cite{touvron2023llama2openfoundation,sennrich2015neural} for tokenizing English text. The original vocabulary size is 32000; we add 768 special tokens for future extensions.

\subsection{Training data collection \& preprocessing}
We train Moonshine on a combination of 90K hours from open ASR datasets and over 100K hours from own internally-prepared dataset, totalling around 200K hours. From open datasets, we use Common Voice 16.1~\cite{commonvoice:2020}, the AMI corpus~\cite{carletta2005ami}, GigaSpeech~\cite{GigaSpeech2021}, LibriSpeech~\cite{panayotov2015librispeech}, the English subset of multilingual LibriSpeech~\cite{Pratap2020MLSAL}, and People's Speech~\cite{galvez2021people}. We then augment this training corpus with data that we collect from openly-available sources on the web. We discuss preparation methods for our self-collected data in the following.


\textbf{Preprocessing noisily-labeled speech.} Many speech sources available on the web have subtitles or captions available, which can serve as labels. However, captions tend be noisy---they may be manually-generated and thus contain text that is orthogonal to the audio content, or they may contain the names of speakers or verbal descriptions of non-speech content. In cases where a manually-generated but possibly-unreliable caption is available, we use a heuristic process to filter out low-quality instances. First, we lowercase and normalize the caption text, removing or replacing, e.g., ambiguous unicode characters, emoji, and punctuation. We then use Whisper large v3 to generate a pseudo-label of the audio content, applying the same text normalization to this pseudo-label as we do the caption. Finally, we compute a normalized Levenshtein distance (between $[0.0, 1.0]$, where $0.0$ is identical and $1.0$ is orthogonal) between the normalized caption and the pseudo-label, filtering out labels with a distance above a threshold. This allows us to treat the human-generated labels in captions as ground truth without introducing excessive noise. After filtering out noisy labels, we prepare the remaining text by applying standardized punctuation and capitalization.

\textbf{Preprocessing unlabeled speech.} The majority of speech available on the web is unlabeled. In these cases, we leverage the Whisper large v3 model to generate training labels for our lighter-weight Moonshine model. The risk inherent in training one model on another model's outputs is that the new model learns the old model's errors. From inspection, we noted that the majority of hallucinated outputs from Whisper large v3 occurred below a predictable value of the average log probability of the output. We thus mitigate the risk of introducing hallucination and other noise in the training set by filtering out instances with an average log probability below this threshold. During this process, we benefited from speed-ups provided by batched inference in the WhisperX implementation~\cite{bain2022whisperx}.

\begin{figure}[]
    \centering
    \includegraphics[width=\columnwidth]{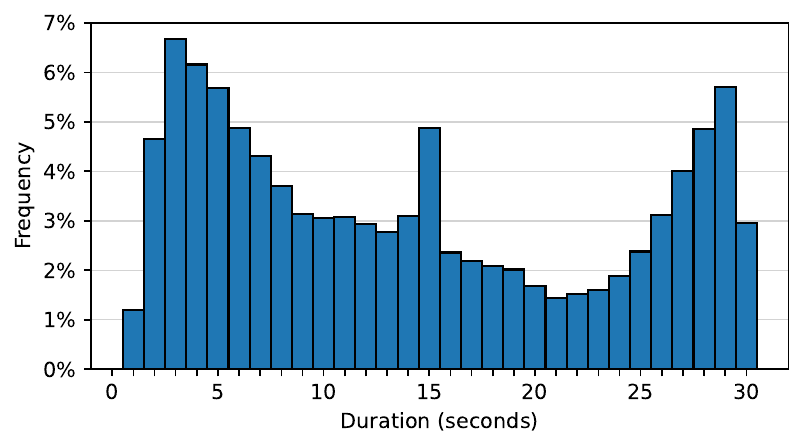}
    \vspace{-3ex}
    \caption{Distribution of training instance durations after combining open and internally-prepared datasets. A slightly bimodal distribution results from our preprocessing procedure, which assembles successive audio segments into instances between 4 and 30 seconds in length.}
    \label{fig:data-histogram}
\end{figure}

\textbf{Controlling instance duration.} We assemble successive speech segments into longer training instances, such that the resulting instance duration is $\in [4, 30]$ seconds with no more than $2$ seconds between successive segments. For our manually-captioned data sources, we use timestamped segments provided by a subtitle file (e.g., an \verb|.srt| file); for our pseudo-labeled audio, we use the timestamps output by Whisper.  Upon combining this data with open datasets, the duration of instances in our aggregate training set obeys a slightly bimodal distribution as depicted in Figure~\ref{fig:data-histogram}.

\subsection{Training}
We performed training on a 32x H100 GPU cluster, using GPU data parallelism provided by Huggingface's Accelerate library~\cite{accelerate}. We used Accelerate's BF16 mixed precision optimizations. We trained the models for 250K steps with a batch size of 32 per GPU (1024 global batch size). We leveraged the AdamW variation of the schedule-free optimizer~\cite{defazio2024roadscheduled}; gradient norm clipping and the initial learning rate warmed up to 1.4e-3 after 8192 steps.

%% file: sections/evaluation.tex
\begin{figure*}[t!]
  \centering
  \includegraphics[width=\textwidth]{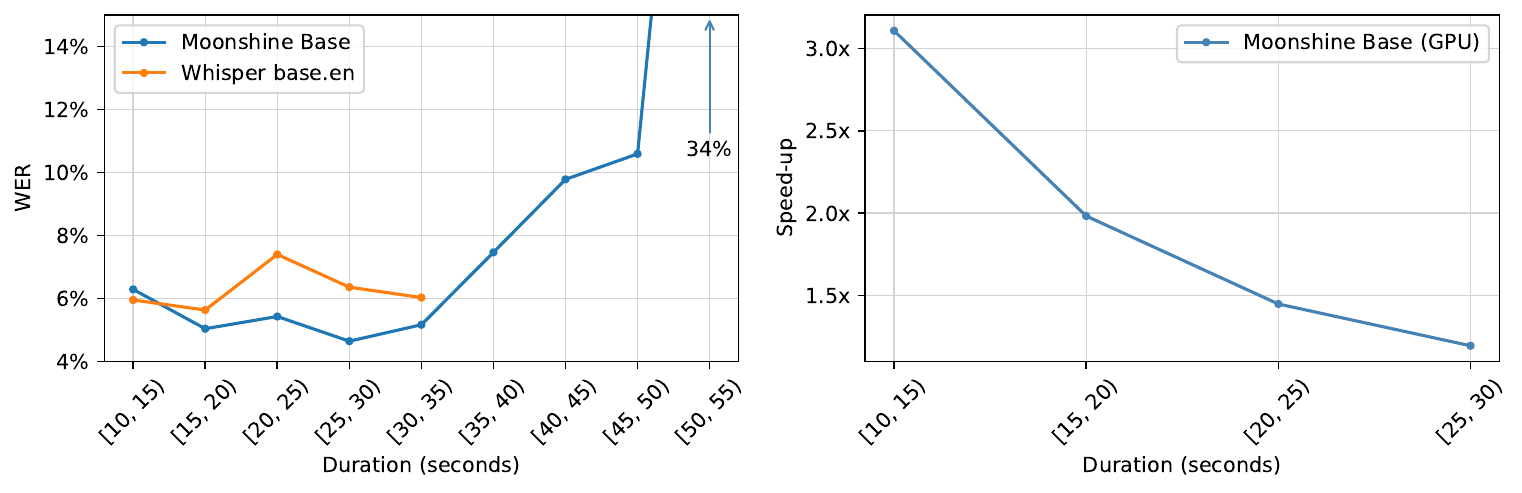}
  \vspace{-3ex}
  \caption{Left: Word Error Rates (WER) across various ranges of input audio
    duration. Right: Speed-up in decoding time of Moonshine Base over
    Whisper \texttt{base.en} across various ranges of input audio duration.}
  \label{fig:wer_rtfx_vs_audio_length}
\end{figure*}

Our experiments measure Moonshine against OpenAI's Whisper models, specifically Whisper \verb|tiny.en| and \verb|base.en|. These two are the most suitable for running on low cost edge devices given their lower memory and compute requirements. We measure the WER on datasets in the OpenASR Leaderboard~\cite{open-asr-leaderboard}. We use greedy decoding, with a heuristic limit of 6 output tokens per second of audio to avoid repeated output sequences.

{\renewcommand{\arraystretch}{1.2}
\begin{table}
  \centering
  \begin{tabularx}{\columnwidth}{| c | C | C |}
      \cline{2-3}
      \multicolumn{1}{c|}{} & Whisper \texttt{base.en} & Moonshine Base \\
      \hhline{|-|-|-|}
      AMI & 21.13 & \cellcolor{green!10}17.79 \\
      \hhline{|-|-|-|}
      Earnings22 & \cellcolor{green!10}15.09 & 17.65 \\
      \hhline{|-|-|-|}
      Gigaspeech & 12.83 & \cellcolor{green!10}12.19 \\
      \hhline{|-|-|-|}
      LibreSpeech (Clean) & 4.25 & \cellcolor{green!10}3.23 \\
      \hhline{|-|-|-|}
      LibreSpeech (Other) & 10.35 & \cellcolor{green!10}8.18 \\
      \hhline{|-|-|-|}
      SPGISpeech & \cellcolor{green!10}4.26 & 5.46 \\
      \hhline{|-|-|-|}
      TEDLium & \cellcolor{green!10}4.87  & 5.22 \\
      \hhline{|-|-|-|}
      Voxpopuli & \cellcolor{green!10}9.76 & 10.81 \\ \hline \hline
      \hhline{|-|-|-|}
      Average & 10.32 & \cellcolor{green!10}\textbf{10.07} \\ \hline
    
  \end{tabularx}
  \caption{WER comparison of base Moonshine vs. Whisper models}
  \label{tab:wer-comparison-base}
\end{table}
}

{\renewcommand{\arraystretch}{1.2}
\begin{table}
  \centering
  \begin{tabularx}{\columnwidth}{| c | C | C |}
      \cline{2-3}
      \multicolumn{1}{c|}{} & Whisper \texttt{tiny.en} & Moonshine Tiny \\
      \hhline{|-|-|-|}
      AMI & 24.24 & \cellcolor{green!10}22.77 \\
      \hhline{|-|-|-|}
      Earnings22 & \cellcolor{green!10}19.12 & 21.25 \\
      \hhline{|-|-|-|}
      Gigaspeech & \cellcolor{green!10}14.08 & 14.41 \\
      \hhline{|-|-|-|}
      LibreSpeech (Clean) & 5.66 & \cellcolor{green!10}4.52 \\
      \hhline{|-|-|-|}
      LibreSpeech (Other) & 15.45 & \cellcolor{green!10}11.71 \\
      \hhline{|-|-|-|}
      SPGISpeech & \cellcolor{green!10}5.93 & 7.7 \\
      \hhline{|-|-|-|}
      TEDLium & 5.97 & \cellcolor{green!10}5.64 \\
      \hhline{|-|-|-|}
      Voxpopuli & \cellcolor{green!10}12.00 & 13.27 \\ \hline \hline
      \hhline{|-|-|-|}
      Average & 12.81 & \cellcolor{green!10}\textbf{12.66} \\ \hline
  \end{tabularx}
  \caption{WER comparison of tiny Moonshine vs. Whisper models}
  \label{tab:wer-comparison-tiny}
\end{table}
}

Moonshine Tiny and Base models achieve better average WER compared to their Whisper counterparts (\verb|tiny.en| and \verb|base.en|, respectively), as depicted in Table~\ref{tab:wer-comparison-base} and Table~\ref{tab:wer-comparison-tiny}. Earnings22 is the dataset for which Moonshine performs the worst compared to Whisper. 8\% of the examples in Earnings22 are less than 1 second long, and the corresponding transcriptions are short utterances such as ``So.'', ``Yes.'', ``Okay.'', etc. Moonshine models tend to produce repeated tokens in the output for such examples, leading to a WER greater than 100\%. Fewer than 0.5\% of examples in our training set are shorter than a second. We hypothesize that this weakens generalization to the Earnings22 dataset.

\begin{figure*}[t!]
  \centering
  \includegraphics[width=\textwidth]{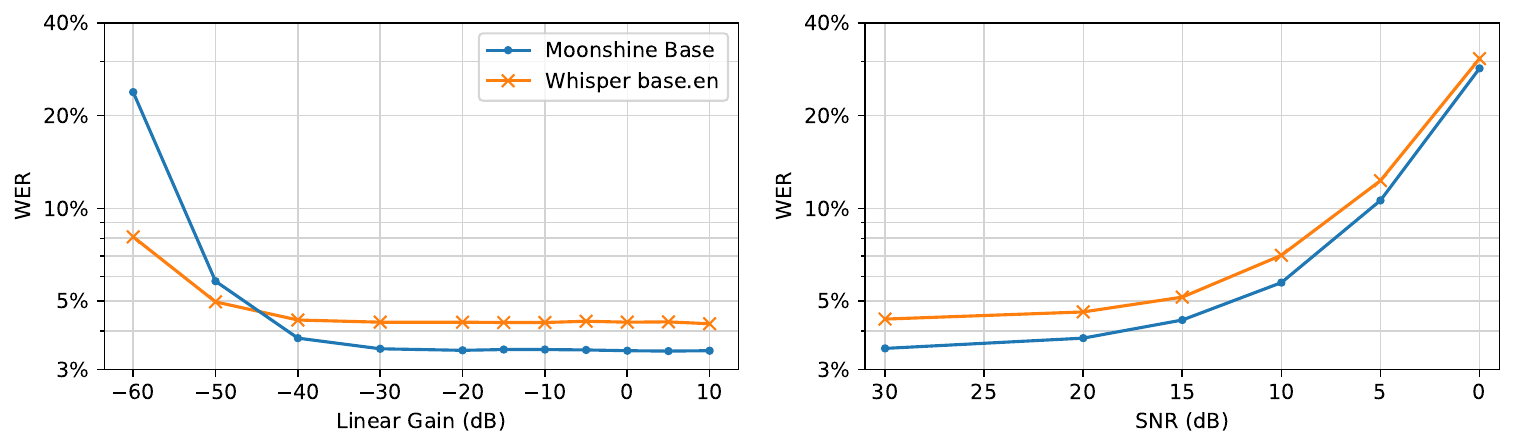}
  \vspace{-2ex}
  \caption{Robustness comparisons on LibriSpeech test-clean. Left: WER as linear gain of test audio increases. Moonshine degrades more at the lowest gains, but maintains superior WER values at -40 dB and beyond. Right: WER as signal-to-noise ratio increases under additive computer fan noise. Moonshine degrades similarly to OpenAI's Whisper counterpart while maintaining superior WER values.}
  \label{fig:wer_snr_additive_noise}
\end{figure*}

\subsection{Accuracy vs. Input Length}

Moonshine models demonstrate the ability to transcribe audio clips of varying lengths. However, we hypothesize that shorter inputs may result in higher WER due to the lack of sufficient contextual information. Conversely, for longer clips---particularly those exceeding the sequence lengths observed during training---there is a notable increase in transcription errors due to hallucinations. To evaluate this behavior, we measured the WER across a range of audio segments from the TEDLium dataset, with clip durations varying between 10 and 55 seconds.

To evaluate inference speed across different clip lengths, we compare Moonshine and Whisper on GPU (H100). The results are depicted in Figure~\ref{fig:wer_rtfx_vs_audio_length}. The speed-up comparison is limited to clips up to Whisper's 30-second maximum, as longer clips require multiple Whisper runs, introducing overhead and skewing the results unfairly against Whisper. Moonshine demonstrates superior performance, particularly on shorter sequences, due to its support for variable sequence lengths in the encoder.

\subsection{Robustness to Input Speech Signal Level}

During this work, we observed that the robustness of Whisper models varies relative to the level of the input speech signal. To measure the impact of input levels (i.e., linear gain) on Moonshine, we tested model robustness as increasing gain is applied to the audio samples in the dataset. As shown in the left portion of Figure~\ref{fig:wer_snr_additive_noise}, Moonshine Base maintains superior WER over most of this range until the input speech signal is very attenuated. At this point (where gain is below -40 dB, i.e., the input audio is very quiet) the WER rises above the Whisper model. For practical applications where high speech attenuation can be avoided in system design, Moonshine Base is robust to varying input levels.

\subsection{Robustness to Additive Noise}

We tested the noise robustness of the Moonshine Base model by measuring the WER for the fan noise observed in a tablet computer application under load. In user studies, we quantified the signal-to-noise ratio (SNR) for this application in the range [9, 17] dB depending on the speaker. We calculate the level of additive noise corresponding to a given SNR based on the average signal power of individual dataset examples with quiet sections removed. The right portion of Figure~\ref{fig:wer_snr_additive_noise} shows how the performance degrades as fan noise increases. Whisper's robustness to noise is known~\cite{radford2022robustspeechrecognitionlargescale}; Moonshine Base maintains this robustness while providing a superior WER.

%% file: sections/conclusion.tex
We briefly discuss limitations before concluding the paper in this section.

\textbf{Architectures and optimizers.} As in any research of this kind, many variations of model architecture could be explored. Likewise, recent advancements in optimizers---particularly Shampoo~\cite{gupta2018shampoo} and SOAP~\cite{vyas2024soap}---show promise for improving WERs in our architecture. Conducting an ablation study on these model architectures and training methods would significantly enhance the community's understanding of the models' limitations. However, due to resource constraints, particularly limited affordability of GPUs, such studies are beyond the scope of this paper. Consequently, we chose our architecture and optimizer based on the authors' experience and a thorough review of relevant literature.

\textbf{Generalization to shorter audio segments.} The relatively low performance that we observed on the Earnings22 dataset motivates collection of more data representative of this case, and investigation of training techniques to introduce more contextual information in the model. This will likely improve model performance with no changes to architecture.

In this paper we introduced Moonshine, a family of lightweight ASR models optimized for low latency, on-device speech-to-text applications. We outlined our model architecture, data collection and preprocessing procedures, and training. We benchmarked Moonshine against two OpenAI Whisper models---\verb|tiny.en| and \verb|base.en|---showing that their Moonshine counterparts provide up to 3x reductions in latency scaled to the duration of input audio. Our work opens the door for new applications of real-time ASR in live transcription, accessibility technologies, and smart devices.

%% file: sections/ack.tex
We would like to acknowledge the developers of the excellent \verb|x-transformers|~\footnote{\href{https://github.com/lucidrains/x-transformers}{https://github.com/lucidrains/x-transformers}} library, which formed the basis of the Moonshine training code. We would also like to thank the folks at Lambda Labs for providing reliable server uptime and alacritous support. Finally, we are thankful for the giants, whose shoulders we all stand on.